\documentclass{PoSLike}

\title{Beam Tests of a Multilayer LumiCal Prototype}

\ShortTitle{Beam Tests of a Multilayer LumiCal Prototype}

\author{\speaker{Oleksandr Borysov} \\
        Tel Aviv University (IL)\\
        E-mail: \email{oleksandr.borysov@cern.ch}}

\author{On behalf of the FCAL collaboration}

\abstract{LumiCal is a sampling electromagnetic calorimeter designed for the precise measurement of integrated luminosity in electron positron linear collider experiments. The present report contains a description and results of the first beam test of a multilayer LumiCal prototype with four silicon detector planes. A 5 GeV electron beam from the CERN PS T9 facility was used to study the performance of the LumiCal prototype. Presented results are mainly focused on the transverse structure of the observed electromagnetic shower and the Moli\`ere radius measurement. A comparison with MC simulation is also discussed.}

\FullConference{Talk presented at the International Workshop on Future Linear Colliders (LCWS2016),\\
                  Morioka, Japan,\\
                  5-9 December 2016.\\
                  C16-12-05.4.}

\begin{document}

\section{Introduction}
\label{Introduction}
Many important open questions of elementary particle physics can be addressed in future e$^{+}$e$^{-}$ \mbox{colliders \cite{LCC_Phys, ILC_Scenarios}}. At present, two such colliders, distinguished by their acceleration techniques and energy reach, are being studied, the International Linear \mbox{Collider (ILC) \cite{ILC_TDR_v1_phys}}, based on superconducting cavities, and the Compact Linear \mbox{Collider (CLIC) \cite{CLIC_CDR, CLIC_UPDATE_YP}} with the two-beam accelerating concept. Two types of detectors are under design for the ILC, the International Large \mbox{Detector (ILD)} and the Silicon \mbox{Detector (SiD) \cite{ILC_TDR_v4_det}}. Similar concepts were developed for CLIC, though a recent study is focused on a single detector model optimized for a 3~TeV centre-of-mass beam energy. Forward regions of the ILC and CLIC detectors are equipped with compact calorimeters designed for the accurate instant and integrated luminosity measurements and for extending the capabilities of the experiments for physics study in the high rapidity region. 
\begin{figure}[h!]
  \begin{minipage}[c]{0.50\textwidth}
    \includegraphics[width=\textwidth]{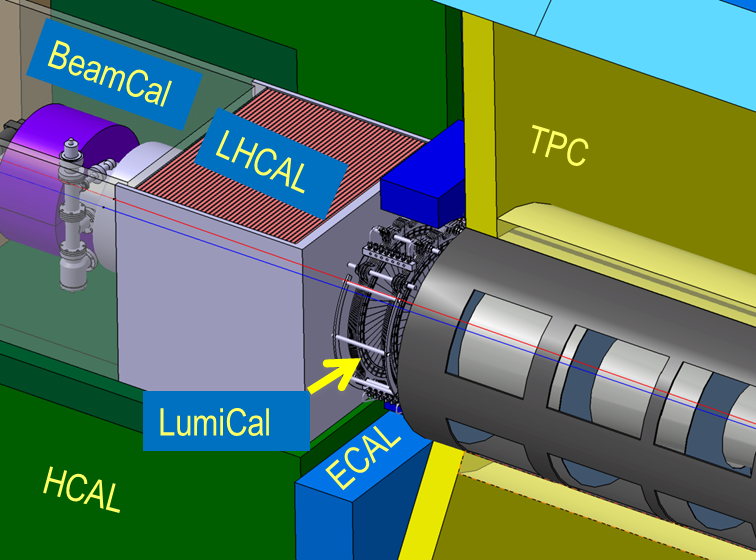}
  \end{minipage}\hfill
  \begin{minipage}[t]{0.45\textwidth}
    \caption{The very forward region of the ILD detector. LumiCal, BeamCal and LHCAL are carried by the support tube 
             for the final focusing quadrupole QD0 and the beam-pipe. TPC, ECAL and HCAL are barrel detectors.
    } \label{fig_forward_ild}
  \end{minipage}
\end{figure}
The layout of one arm of the forward region of the ILD experiment is presented in~Fig.~\ref{fig_forward_ild}. LumiCal in ILD design is an electromagnetic sampling calorimeter with 30 layers of 3.5~mm (1X$_0$) thick tungsten absorbers and silicon sensors placed in one millimeter gap between absorber plates. In case of CLIC it has a similar design, but the number of layers is increased up to 40 because of higher beam energy. The LumiCal aims on a precise measurement of the integrated luminosity using Bhabha scattering as a gauge process. Its design was developed and optimized in MC simulation studies~\cite{FCAL_ILC}. Several beam tests were carried out in the past to study and verify the performance of a single LumiCal detector module~\cite{TB2010_jinst}. It allowed to develop the procedure of the signal processing thus achieving a signal to noise ratio in the range of 19-23 after the common mode correction. 

As the next step in calorimeter development, the same detector modules are assembled in a stack as sensitive layers between tungsten absorbers. The beam test of such assembly can demonstrate the performance of the detector modules in multilayer configuration. Specifically it aims to measure the signal to noise ratio, the level of common mode noise and to test the electromagnetic shower development. Of particular importance is an investigation of the position reconstruction which corresponds to the polar angle measurement in the LumiCal as a luminosity detector in e$^{+}$e$^{-}$ colliders. The detailed description of the test beam setup and some results were presented earlier~\cite{TB2014_Whistler_proc, TB2014_ICHEP2016}. This work presents the recent results of the data analysis and MC simulations of the first multilayer LumiCal prototype tested with 5~GeV electron beam.

\section{LumiCal sensor and test beam setup}
\label{TB_Setup}
\begin{figure}[h!]
  \centering
  \includegraphics[width=0.9\textwidth]{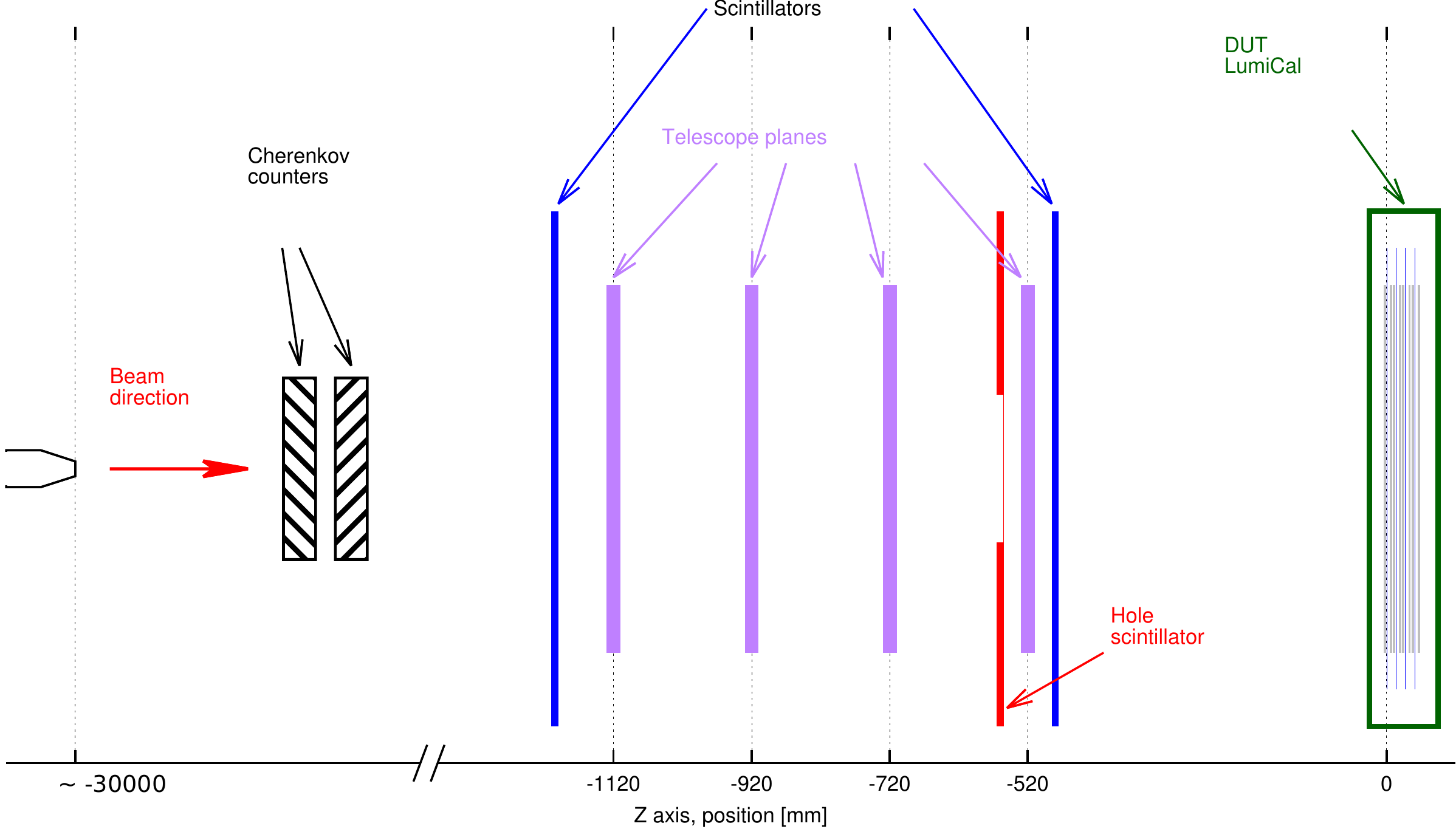}
  \caption{Test beam area instrumentation geometry. Not in scale.
  } \label{fig_total_geom}
\end{figure}

The test beam was performed at the PS east area test beam facility T9 which provides a secondary beam of muons, pions, hadrons and electrons with momenta in the range of 1-15~GeV/c. A narrow band of particle momenta centered at 5 GeV is selected using a dipole magnetic field and a set of collimators. The Cherenkov counters are used to provide a trigger for the electrons or/and muons. The schematic diagram of the instrumentation geometry is shown in Fig.~\ref{fig_total_geom}. Four planes of pixel detectors, so-called telescope, are set upstream of the calorimeter to enable position reconstruction of incoming beam particles. Each pixel detector contains one MIMOSA-26 chip~\cite{MIMOSA_26} with active an area of 21.2$\times$10.6~mm$^{2}$. The position resolution achieved with the telescope is around 9~$\mu$m for each coordinate. 

Three scintillation counters are used to provide a trigger for particles traversing the active part of the telescope sensors and the active region of the calorimeter. Two 5$\times$5~cm$^{2}$ scintillator tiles are placed upstream and downstream of the telescope (marked in blue in Fig.~\ref{fig_total_geom}) and one (marked in red), with a 9~mm diameter circular hole, is placed just before the last telescope plane. In order to ensure that triggers are only generated by beam particles in the sensitive area of the telescope, the signal from the hole scintillator is set in anti-coincidence.  The trigger signal is combined with the
Cherenkov counters response to create the trigger for electrons and/or muons.

\begin{figure}[h!]
  \begin{minipage}[c]{0.55\textwidth}
    \includegraphics[width=\textwidth]{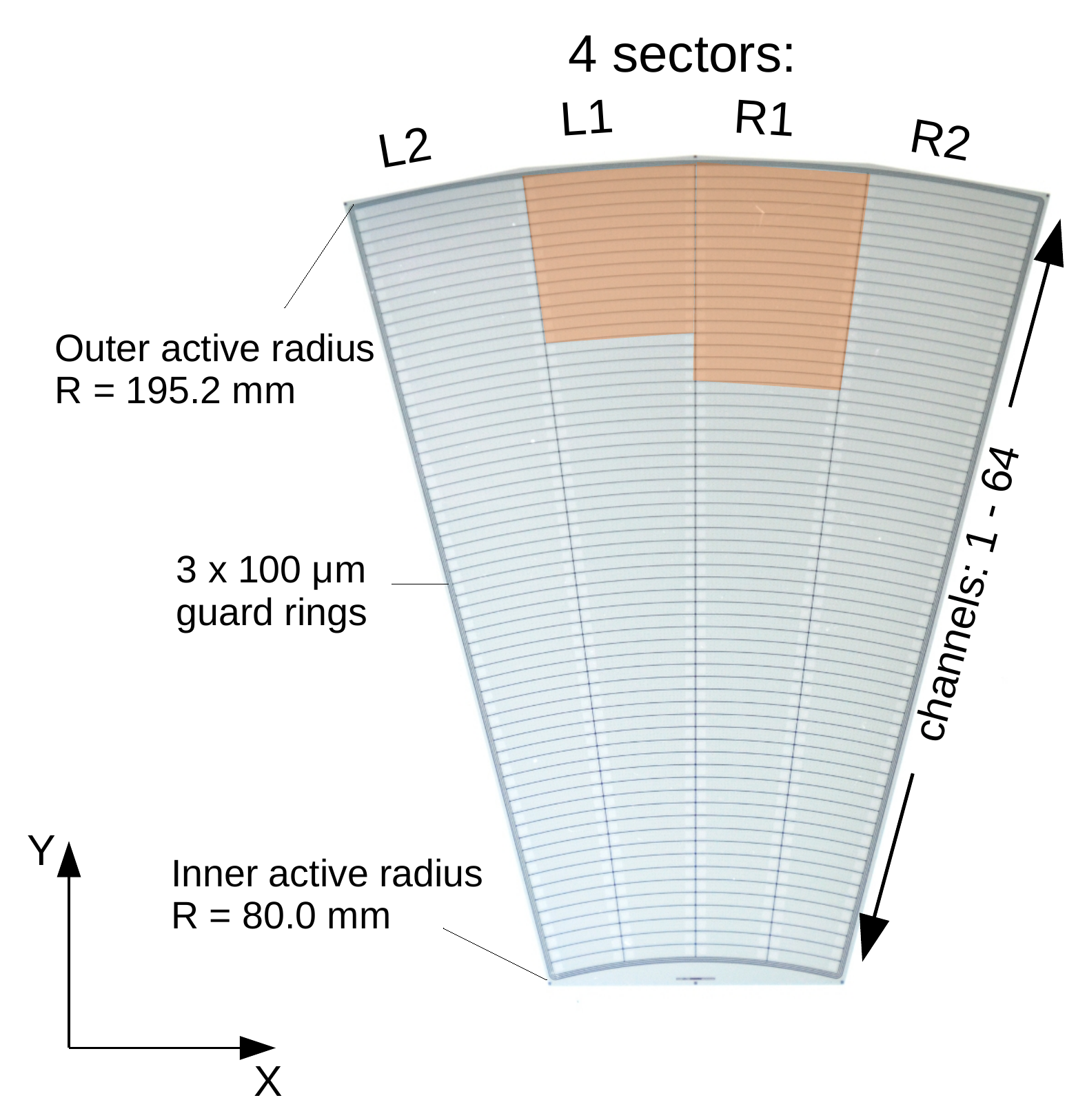}
  \end{minipage}\hfill
  \begin{minipage}[t]{0.42\textwidth}
    \caption{A LumiCal silicon pad sensor. Highlighted area in the top part of sectors L1 and R1 marks the pads which are connected to read-out electronics.}
    \label{fig_si_sensor}
  \end{minipage}
\end{figure}

Four LumiCal electronic modules were assembled for this test beam. Each module consists of one LumiCal silicon sensor glued to the supporting PCB and front-end electronics assembled on a separate PCB. The silicon sensor is made of a 320~$\mu$m thick high resistivity n-type silicon wafer. It has a shape of a circular sector with a central angle of 30$^{\circ}$, with inner and outer radii of the sensitive area of 80~mm and 195.2~mm, respectively~(see Fig.~\ref{fig_si_sensor}). 256 pads made as p-type implants are arranged in four sectors with 64 pads of 1.8~mm pitch in each sector. In total, 32 pads were connected to the readout electronics: 14 in sector L1 and 18 in sector R1 as shown in Fig.~\ref{fig_si_sensor}. 

In order to study the development of the electromagnetic shower, three configurations of absorbers and sensitive layers are used. It allows to sample the electromagnetic shower at 1X$_0$ and then from 3X$_0$ to 9X$_0$ with a step of 1X$_0$, assuming that each tungsten plate corresponds to one radiation length (1X$_0$) which is approximately correct.

\section{Position reconstruction}
\label{TB_POsition_Reco}
Different methods were elaborated for the electron position reconstruction in electromagnetic calorimeters~\cite{Position_reco_1, Position_reco_algs, Position_reco_logw}. Most of them assume that lateral shower development has single or double exponential profiles. A rather simple and efficient method was described in~\cite{Position_reco_logw}. It suggests to use a center of gravity formula, but instead of weights which are linear in the deposited energy, the weights to be used are given by the following expression ("logarithmic weighting"):
\begin{equation} \label{eq:log_weight}
w_n = max\left\{0; W_0+\ln{\frac{E_n}{\sum\limits_{n} E_n}}\right\},
\end{equation}
where $E_n$ is the energy deposited in pad $n$ and $W_0$ is a free dimensionless parameter. Its value can be optimized for a given calorimeter to provide the best position resolution. 
It was shown~\cite{Position_reco_algs} that particle position reconstruction in electromagnetic calorimeter from the fit to the lateral shower profile gives good resolution and has a relatively small bias depending on the particle position withing the sensor segmentation unit. It has the advantage that it can be used without apriori knowledge of shower characteristics. It was also demonstrated that the core of the shower has dominant contribution to the chi-squared of the fitting method. In this way the method counteracts the event-by-event fluctuations in the lateral shower development.

The design of the LumiCal sensor is optimized for the polar angle measurement in collider experiment. It has a fine pitch in the radial direction which corresponds to $y$ direction in Fig.~\ref{fig_si_sensor} and a relatively large pad size in the azimuthal direction. That is why it is mainly interesting to study position reconstruction with respect to the radial direction of the sensor. 

\begin{figure}[h!]
  \begin{minipage}[t]{0.47\textwidth}
    \includegraphics[width=\textwidth]{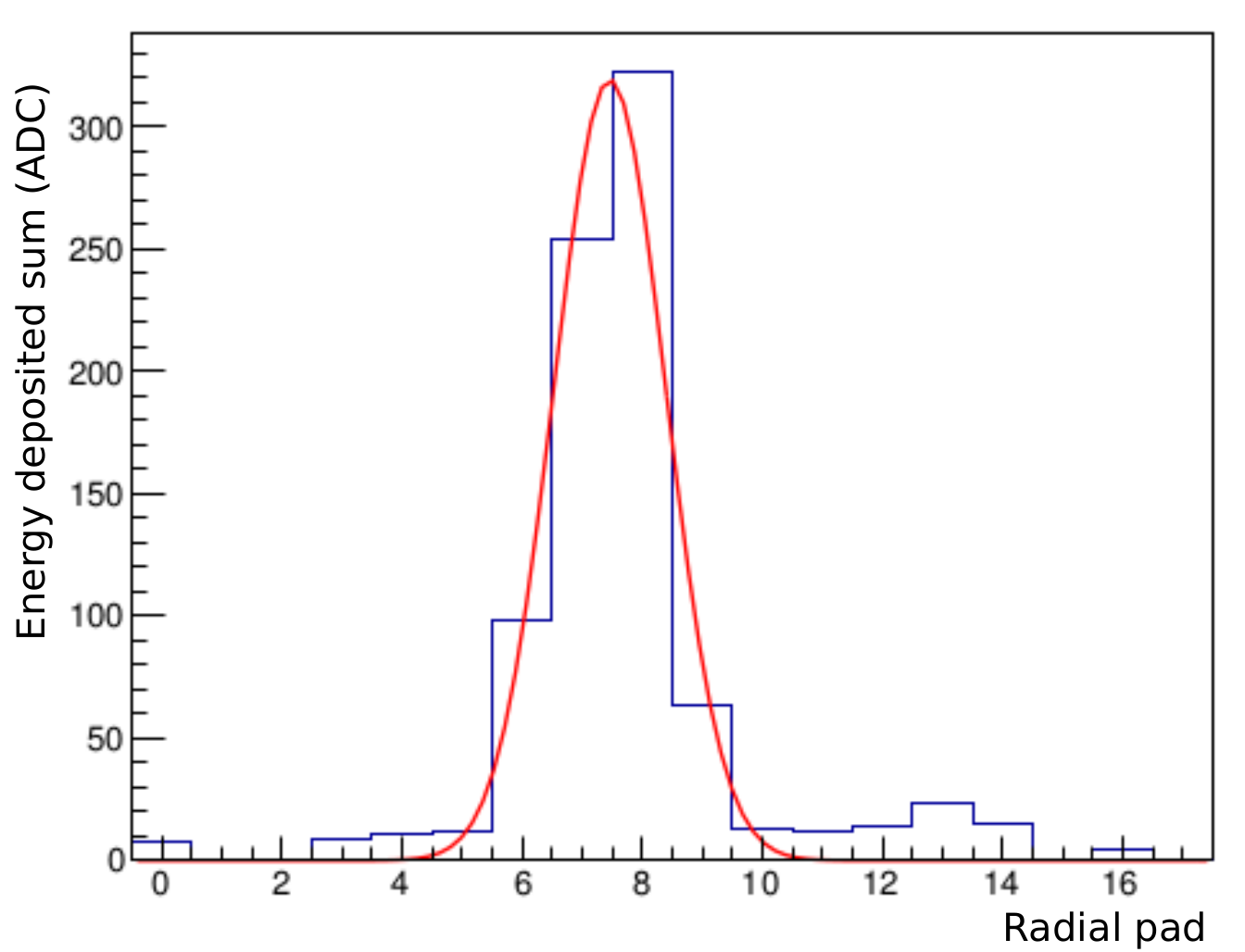}
    \caption{Deposited energy in a single event observed in the LumiCal prototype. The red line is a Gaussian fit to determine the shower $y$ position (one unit of pad is equivalent to 1.8~mm).}
    \label{fig_event_gaus_fit}
  \end{minipage}\hfill
  \hspace{0.06\textwidth}
  \begin{minipage}[t]{0.4\textwidth}
    \includegraphics[width=\textwidth]{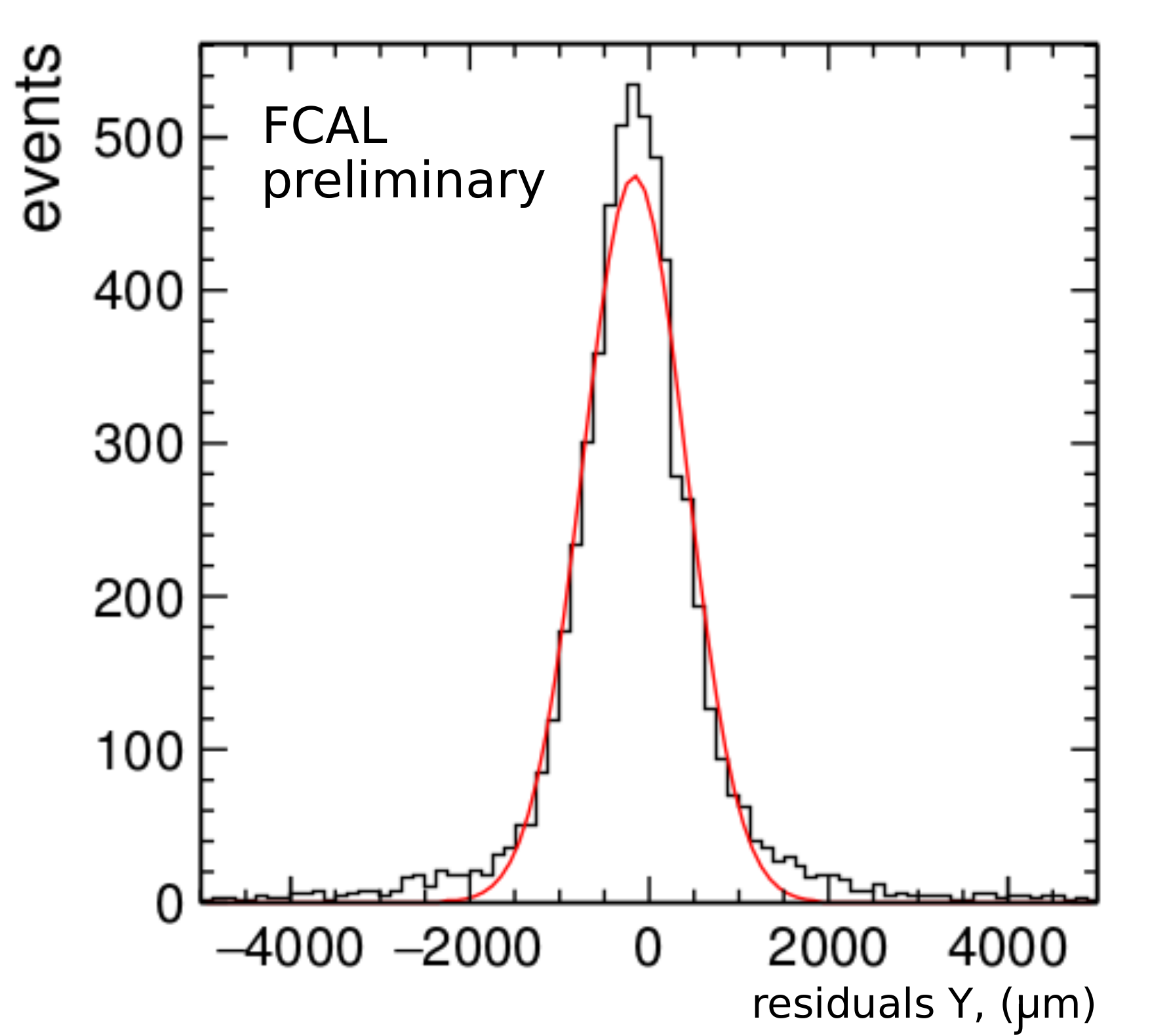}
    \caption{Distribution of residuals between reconstructed shower $y$ position and the one projected from the telescope. Red line is Gaussian fit, estimated resolution is 0.5~mm.}
    \label{fig_y_resolution}
  \end{minipage}\hfill
\end{figure}

Let denote by $\varepsilon_{nkl}$ the energy deposited in the sensor pad for the layer~$l$, sector number~$k$ and radial pad index~$n$, then the one dimensional lateral deposited energy distribution for one event is constructed using the following sum:
\begin{equation} \label{eq:E_rad}
E_n = \sum\limits_{k,l} \varepsilon_{nkl} \ ,
\end{equation}
where the layer index~$l$ runs over all sensitive planes in the stack from 1 to 4 and the sector index $k$ for two sectors according to the part of the sensors connected to readout electronics. An example of the~$E_n$ distribution for a single event is presented in~Fig~\ref{fig_event_gaus_fit}.

To test the position reconstruction in the LumiCal prototype, the fitting method was used. It requires a choice of the trial function, which will be discussed in details in section~\ref{TB_Shower}. The important note is that position reconstruction is mainly defined by the shower core. It also follows from the logarithmic weighting method, where the value of $W_0$ impose the minimum threshold on the energy deposited in the sensor pads which are considered in position calculation.
Fig.~\ref{fig_event_gaus_fit} shows the lateral profile of the shower in a single event observed in the LumiCal prototype fitted with a Gaussian function. The resolution estimated from the distribution of residuals between the position reconstructed in the calorimeter and the one projected from the telescope (Fig.~\ref{fig_y_resolution}) is about 0.5~mm.   	
%

\section{Electromagnetic shower in LumiCal prototype}
\label{TB_Shower}

The recent analysis of the test-beam data is devoted to the study of the transverse structure of the electromagnetic shower and Moli\`ere radius measurement. The average distribution of the deposited energy in the transverse plane is symmetric with respect to the longitudinal shower axis and does not depend on the azimuthal angle. Its radial dependence is characterized by a narrow core, and a broadening tail. Several approaches were used to approximate this distribution: weighted sum of two exponential functions~\cite{Position_reco_1, double_exp_w-si1}, two Gaussians~\cite{pdg}. Another function was used by Grindhammer and Peters~\cite{Grindhammer1}. A function similar to the last one is used for the tail description in the present study while the core is approximated by a Gaussian:
 \begin{equation} \label{eq:MR_FrFinel}
F ( r ) =  (A_C)e^{-(\frac{r}{R_C})^2} + ( A_T )\frac{2rR_T^2}{ (r^2 + R_T^2 )^2 } \ .
\end{equation}
Here $R_C$ ($R_T$ ) is the median of the core (tail) component and $A_C$, $A_T$ are their weights. On average, only 10\% of the deposited energy lies outside a cylinder with a radius $R_\mathcal{M}$. This property can be used in order to estimate the size of the electromagnetic shower in the lateral direction. Assuming the function~(\ref{eq:MR_FrFinel}) is normalized to unity, the value of $R_\mathcal{M}$ can be found from the following equation: 
\begin{equation} \label{eq:MR_1}
0.9 = \int_{0}^{2\pi } d\varphi \int_{0}^{R_\mathcal{M}}F ( r ) r dr \ ,
\end{equation}
The function $F(r)$ can be reconstructed using the test-beam data or the MC simulation. 

The design of the LumiCal sensor with fine granularity in one direction ($y$) allows detailed sampling of the transverse structure of the shower only in this $y$ direction. The measured distribution~$E_n$ expressed by the formula~\ref{eq:E_rad} corresponds to $F(r)$ integrated over the area of the pads with the same radial position in the LumiCal sensor. Since the area of the LumiCal sensor connected to the read-out electronics corresponds to the big radii, it is a good approximation to treat the pads as straight strips. In this approach the lateral distribution~$E_n$ can be approximated by integrating~$F(r)$ along the Cartesian coordinate $x$:
\begin{equation} \label{eq:math_edep_1d}
G(y) =\int_{X_{min}}^{X_{max}} F(\sqrt{x^2+y^2})dx \ . 
\end{equation}
Here, the polar coordinate $r=\sqrt{x^2+y^2}$ is substituted with Cartesian ($x,y$), the range $(X_{min}, X_{max})$ is defined by the sensor geometry which corresponds to two sectors. Depending on the form of the trial function~$F(r)$, the integration can be performed either analytically or numerically. By fitting $G(y)$ to the shower average lateral profile~$E_n$~(~\ref{eq:E_rad}), and finding its parameters, the original $F(r)$ can be recovered and then used to calculate the Moli\`ere~radius $R_\mathcal{M}$ from the equation (\ref{eq:MR_1}).

\begin{figure}[h!]
  \begin{minipage}[t]{0.47\textwidth}
    \includegraphics[width=\textwidth]{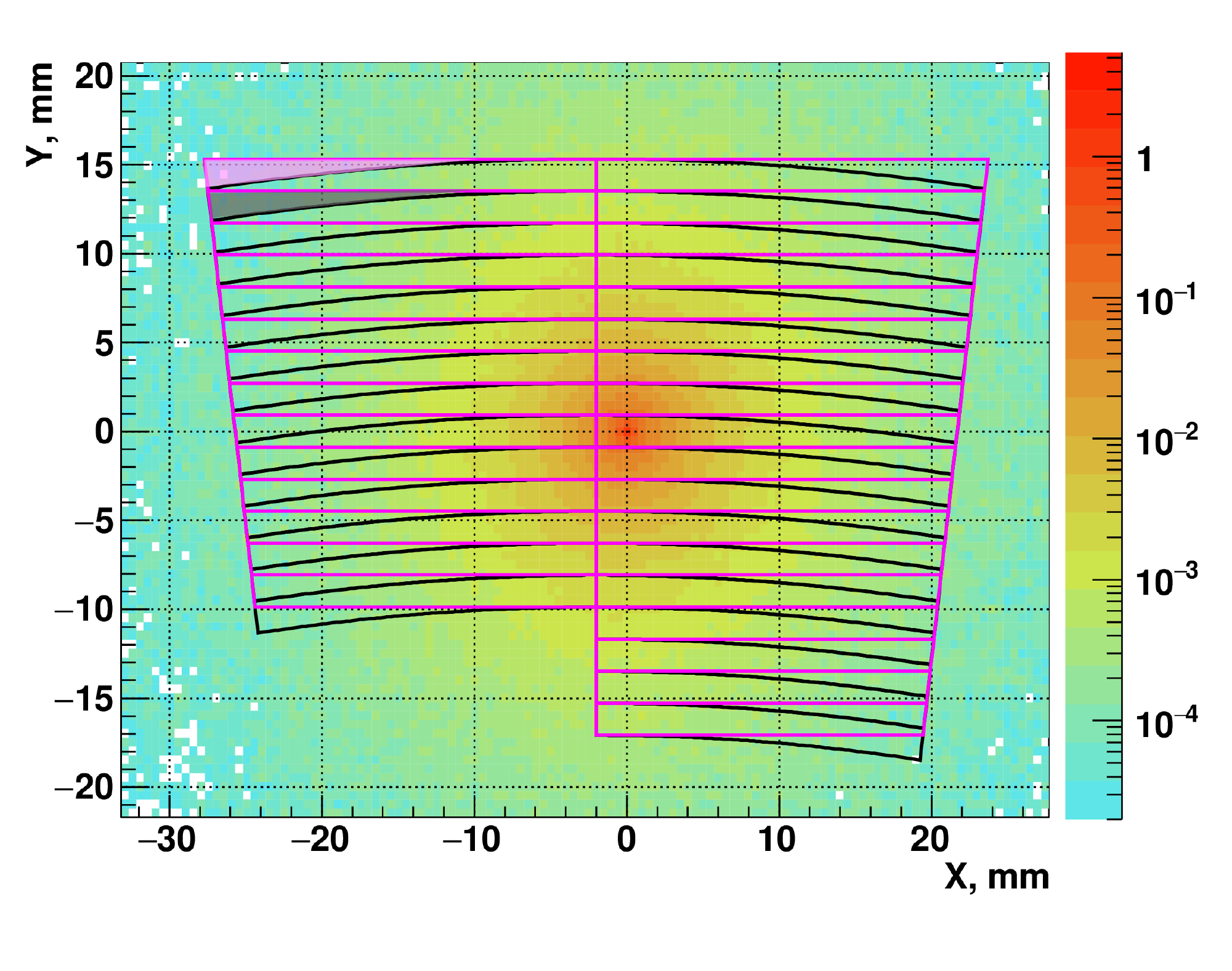}
    \caption{Two dimensional average distribution of the deposited energy in the calorimeter. The black grid represents the sector-like pads of the LumiCal sensor, magenta --- strip-like pads.}
    \label{fig_e_dep_2d_arc_trapez}
  \end{minipage}\hfill
  \hspace{0.05\textwidth}
  \begin{minipage}[t]{0.47\textwidth}
    \includegraphics[width=\textwidth]{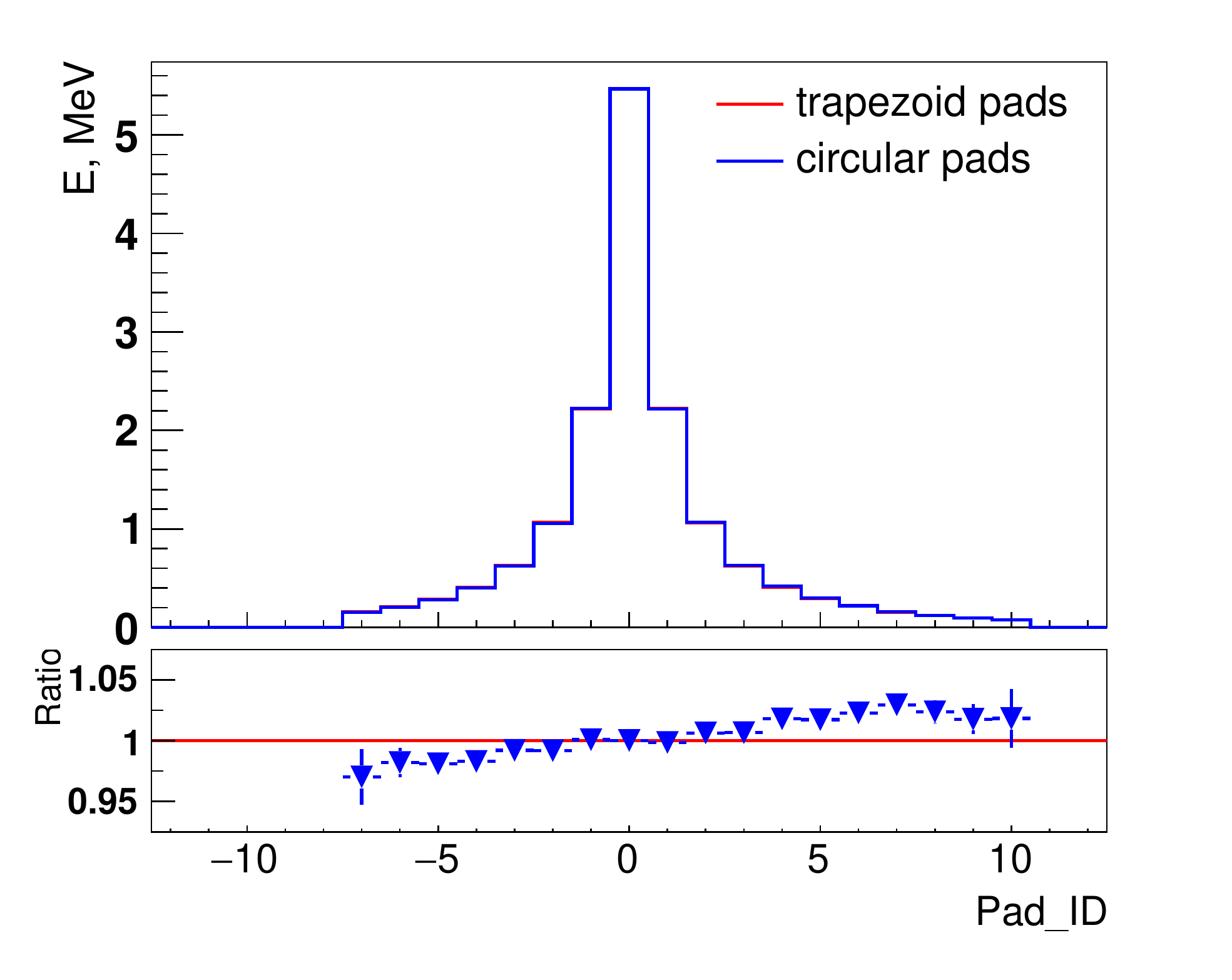}
    \caption{Comparison of 1D lateral profiles of the deposited energy in the calorimeters for different geometries of the sensor pad. Averaging within the beam spot area is performed, (one unit of pad is equivalent to 1.8~mm).}
    \label{fig_e_dep_arc_vs_trapez}
  \end{minipage}\hfill
\end{figure}

An additional MC simulation with slightly modified geometry of the calorimeter is used in order to estimate the possible affect of the above mentioned approximation. For this study the sensitive detector of the sampling calorimeter is implemented with a fine granularity of 0.5$\times$0.5~mm$^2$ and the transverse size of the calorimeter is extended up to 40$\times$40~cm$^2$. Twenty thousand electrons with 5 GeV/c momenta are simulated. The average deposited energy in the transverse plane is presented in Fig.~\ref{fig_e_dep_2d_arc_trapez} as a contour plot. The lateral one dimensional distribution~$E_n$ is obtained by integrating the deposited energy over the sensor pads presented in the figure as grids. The beam spot can be taken into account in this approach by changing the position of the sensor with respect to the 2D distribution. The black grid represents the geometry of the LumiCal sensor, while the magenta one corresponds to the strip-like pads of a trapezoidal shape. The comparison between the one dimensional lateral profile for these two sensor pad geometries is shown in Fig.~\ref{fig_e_dep_arc_vs_trapez}. The lowest four pads of sector R1 are not considered. One can see that the effect becomes noticeable at a level of 2\% for the pads distant from the shower core. The difference in the integration area for one of the pads for different geometries is highlighted by the shadowed triangles. The integrand function, represented by the 2D distribution of deposited energy, is four orders of magnitude smaller in this area than in its maximum and it changes relatively weakly with the distance. It explains the fact that the difference of the integrals is small. There is also a systematic increase in the fraction of deposited energy collected by the LumiCal sensor pad compared to the trapezoid one. It can be seen that the LumiCal pads are bent closer to the center when they are "above" the maximum of the shower and it corresponds to slightly higher values of the integrand. In the same time, if they are "below" the center they are bent away from the shower maximum and collect less energy than straight strips. For a rather wide range of pads close to the shower core the difference is vanishing.

\begin{figure}[!h]
  \centering
  \includegraphics[width=0.6\textwidth]{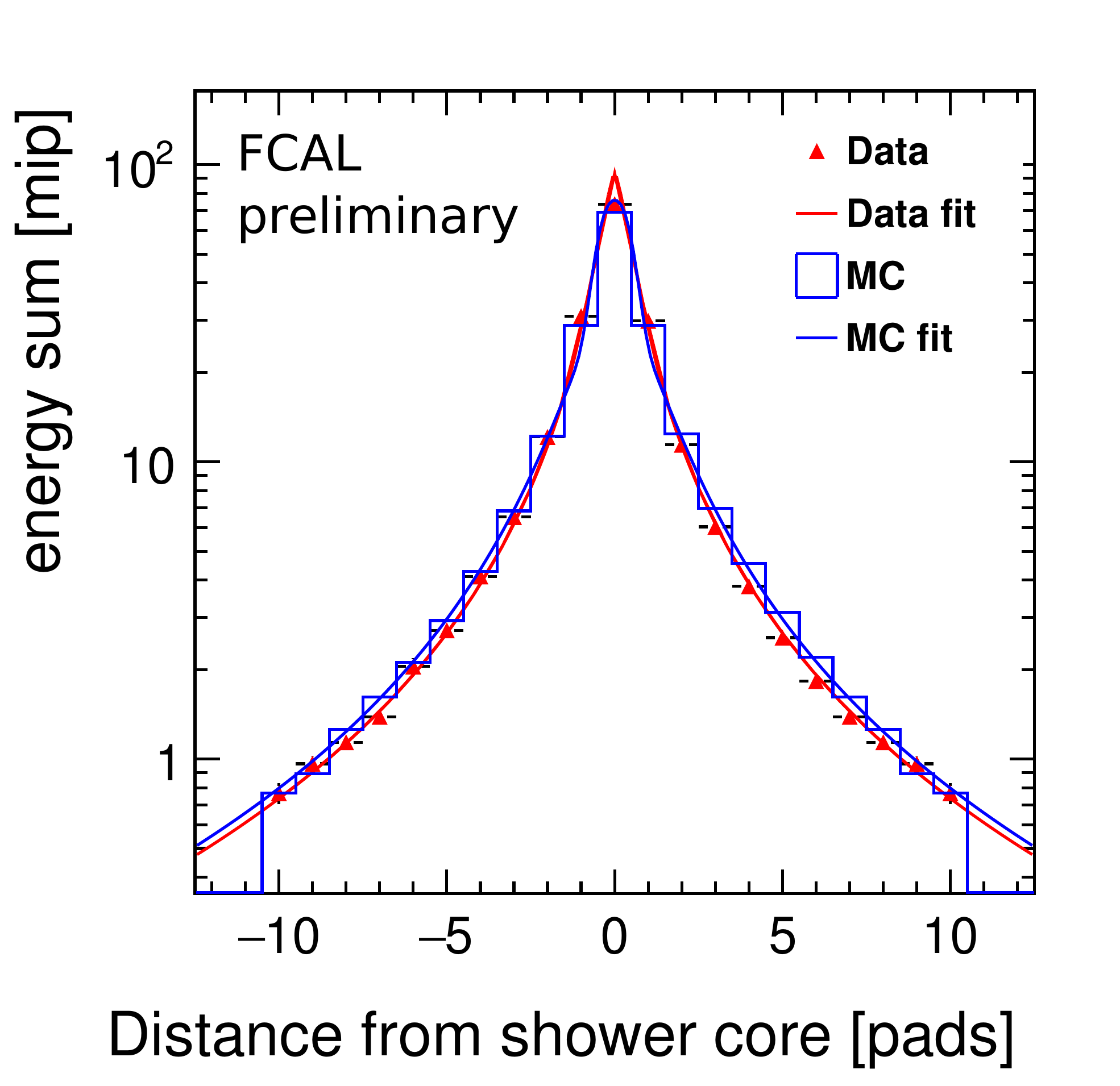}
  \caption{Shower average lateral profile of test-beam data and MC simulation for one of the setup configurations. The line represents the fit of G($y$) to the data (red) and the MC (blue).}
  \label{fig:MC_CONF2_fit}
\end{figure}

After considering different functions used to describe the transverse shower profile, we found that using a Gaussian distribution for the shower core and a rational function similar to~\cite{Grindhammer1} for the shower tail, as shown in equation~(\ref{eq:MR_FrFinel}), gives the best representation to the data and the MC simulation. The fit and the solution of the algebraic equation~(\ref{eq:MR_1}) are made numerically using ROOT package.

The following selection criteria to the events are applied to produce the one dimensional lateral distribution of the deposited energy:
\begin{itemize}
  \item events with only one track, based on telescope reconstruction;
  \item only tracks whose incident position in calorimeter is within 600~$\mu$m of the central band along the pad of the LumiCal sensor.
\end{itemize}
The last requirement allows cleaner events folding without taking into account the dependence of deposited charge distribution between pads on the particle position inside the pad. It reduces statistics, but eliminates the contribution of charge sharing to the final systematic uncertainty. 
 
The average lateral profile of the deposited energy for the data and MC simulation for one of the configurations of the beam test setup together with fit functions are shown in Fig.~\ref{fig:MC_CONF2_fit}. In this configuration four detector modules are placed after 3, 5, 7 and 9 tungsten absorbers (it roughly corresponds to shower sampling at 3X$_0$, 5X$_0$, 7X$_0$ and 9X$_0$). A similar distribution can be constructed combining the results of different beam test setup configurations where some layers are identical and others are complementary. The Moli\`ere radius for such combination can be calculated following the same procedure. The data analisis is currently in progress and more studies are required for better understanding of systematic uncertainty of the Moli\`ere radius measurement.
%
\section{Summary}
\label{Summary}
Recent results of the beam test of the first multi-plane prototype of the electromagnetic calo-rimeter LumiCal have been presented. They are mainly devoted to the study of electromagnetic shower development in the transverse plane using collected data and MC simulation. The method of measuring the Moli\`ere radius adapted to specific segmentation of the LumiCal sensor has been described. The data analisis is in progress focused on the systematic uncertainty evaluation of the Moli\`ere radius measurement.

\section{Acknowledgments}
This work is partly supported by the Israel Science Foundation (ISF), the I-CORE program of the Israel Planning and Budgeting Committee. This project has received funding from the European Union's Horizon 2020 Research and Innovation programme under Grant Agreement no. 654168.



\begin{thebibliography}{99}
   \bibitem{LCC_Phys}
   Keisuke Fujii, Christophe Grojean, Michael E. Peskin, Tim Barklow, Yuanning Gao, Shinya Kanemura, Hyungdo Kim, Jenny List, Mihoko Nojiri, 
   Maxim Perelstein, et al. \emph{The Potential of the ILC for Discovering New Particles}. DESY 17-012 (KEK Preprint 2016-60, 
   SLAC-PUB-16916, LAL 17-017, MPP-2017-5, IFT-UAM/CSIC-17-008). arXiv:1702.05333 [hep-ph].
   
   \bibitem{ILC_Scenarios}
   T. Barklow, J. Brau, K. Fujii, J. Gao, J. List, N. Walker and K. Yokoya, \emph{ILC Operating Scenarios}, arXiv:1506.07830 [hep-ex].
   
   \bibitem{ILC_TDR_v1_phys}
   Howard Baer, Tim Barklow, Keisuke Fujii, Yuanning Gao, Andre Hoang, Shinya Kanemura, Jenny List, Heather E. Logan, Andrei Nomerotski, Maxim Perelstein, 
   Michael E. Peskin, et al. \emph{The International Linear Collider. Technical Design Report, Volume 2: Physics}. 2013. arXiv:1306.6352 [hep-ph].
   
   \bibitem{CLIC_CDR}
   P.~Lebrun, L.~Linssen, A.~Lucaci-Timoce, D.~Schulte, F.~Simon, S.~Stapnes, N.~Toge, H.~Weerts, J.~Wells, \emph{The CLIC Programme: 
   Towards a Staged e+e- Linear Collider Exploring the Terascale: CLIC Conceptual Design Report.} arXiv:1209.2543 [physics.ins-det]
   
   \bibitem{CLIC_UPDATE_YP}
   The CLIC, CLICdp collaborations, \emph{Updated baseline for a staged Compact Linear Collider.} CERN Yellow Report CERN-2016-004;
    arXiv:1608.07537 [physics.acc-ph] 

   \bibitem{ILC_TDR_v4_det}
   \emph{The International Linear Collider. Technical Design Report, Volume 4}: Detectors, 2013. arXiv:1306.6329 [physics.ins-det].
   
   \bibitem{FCAL_ILC}
   H. Abramowicz et al., \emph{Forward instrumentation for ILC detectors.} JINST 5 (2010) P12002.

   \bibitem{TB2010_jinst}
    H. Abramowicz et al., \emph{Performance of fully instrumented detector planes of the forward calorimeter of a
     Linear Collider detector.} JINST 10 (2015) P05009.

   \bibitem{TB2014_Whistler_proc}
    O.~Borysov, V.~Ghenescu, A.~Levy, I.~Levy, S.~Lukic, J.~Moron, A.T.~Neagu, T.~Preda, 
    O.~Rosenblat~(On behalf of the FCAL collaboration), \emph{Results from the October 2014 CERN test beam of LumiCal.} 
    Proceedings of the LCWS2015, Whistler BC Canada, November 2-6, 2015.
    
   \bibitem{TB2014_ICHEP2016}
    O.~Borysov on behalf of the FCAL Collaboration, \emph{Recent progress with very forward calorimeters for linear colliders.}
    Proceedings, 38th International Conference on High Energy Physics (ICHEP 2016): Chicago, IL, USA, August 3-10, 2016. PoS(ICHEP2016)233.

   \bibitem{MIMOSA_26}
    J. Baudot et al., \emph{First test results of MIMOSA-26: A fast CMOS sensor with integrated zero suppression and digitized output},
    http://inspirehep.net/record/842163?ln=en.
    
   \bibitem{Position_reco_1}
   G.~A.~Akopdjanov, A.~V.~Inyakin, V.~A.~Kachanov, R.~N.~Krasnokutsky, A.~A.~Lednev, Yu.~V.~Mikhailov, 
   Yu.~D.~Prokoshkin, E.~A.~Razuvaev and R.~S.~Shuvalov. 
   \emph{Determination of Photon Coordinates in A Hodoscope Cherenkov Spectrometer.} Nucl. Instr. and Meth. 140 (1977) 441.
   
   \bibitem{Position_reco_algs}
   L.~Bugge. \emph{On the determination of shower central positions from lateral samplings.} Nucl. Instr. and Meth. A242 (1986) 228.

   \bibitem{Position_reco_logw}
   T.C.~Awes, F.E.~Obenshain, F.~Plasil, S.~Saini, S.P.~Sorensen, G.R.~Youn. \emph{A simple method of shower 
   localization and identification in laterally segmented calorimeters.} Nucl. Instr. and Meth. A311 (1992) 130.

   \bibitem{double_exp_w-si1}
   G.~Ferri et al., \emph{The Structure of Lateral Electromagnetic Shower Development in Si/W And Si/U Calorimeters.} 
   Nucl. Instr. and Meth. A273 (1988) 123.
   
   \bibitem{pdg}
   K.A. Olive et al., \emph{(Particle Data Group)}, Chin. Phys. C, 38, 090001 (2014) and 2015 update.

   \bibitem{Grindhammer1}
   G.~Grindhammer, M.~Rudowicz, and S.~Peters, 
   \emph{The Parameterized Simulation of Electromagnetic Showers in Homogeneous and Sampling Calorimeters.} 2000, hep-ex/0001020.

\end{thebibliography}
\end{document}